\documentclass[published]{JHEP3} % 10pt is ignored!

\JHEP{00(2007)000}

\JHEPspecialurl{http://jhep.sissa.it/JOURNAL/JHEP3.tar.gz}

\usepackage{epsfig,multicol,bbm}

\newcommand\fverb{\setbox\fverbbox=\hbox\bgroup\verb}
\newcommand\fverbdo{\egroup\medskip\noindent%
			\fbox{\unhbox\fverbbox}\ }
\newcommand\fverbit{\egroup\item[\fbox{\unhbox\fverbbox}]}
\newbox\fverbbox

\title{Lepton asymmetries and the growth 
of cosmological seed magnetic fields}

\author{V.B. Semikoz\thanks{Pushkov Institute of Terrestrial Magnetism,
  Ionosphere and Radiowave Propagation of the Russian Academy of
  Sciences, IZMIRAN, Troitsk, Moscow region, 142190, Russia.}} 
\author{J.~W.~F.~Valle\\
	AHEP Group, Institut de F\'{\i}sica Corpuscular --
 C.S.I.C./Universitat de Val{\`e}ncia 
  Edificio Institutos de Paterna, Apt 22085, E--46071 Valencia, Spain.
	E-mail: \email{semikoz@ific.uv.es},\email{valle@ific.uv.es}}
\received{\today} 		%%
\accepted{\today}		%% These are for published papers.

\preprint{IFIC/07-19}	% OR: \preprint{Aaaa/Mm/Yy\\Aaa-aa/Nnnnnn}
\abstract{Primordial cosmological hypermagnetic fields polarize the
  early Universe plasma prior to the electroweak phase transition
  (EWPT). As a result of the long range parity violating gauge
  interaction present in the Standard Model their magnitude gets
  amplified, opening a new perturbative way of seeding the primordial
  Maxwellian magnetic field at EWPT.}


\begin{document} 

The electroweak phase transition (EWPT) has long been considered as a
playing an important role in the generation of primordial magnetic
fields~\cite{Vachaspati:1991nm,Joyce:1997uy,Grasso:2000wj}. Here we
suggest that the seed fields accounting for the observed intergalactic
magnetic fields arise, say, as tiny fluctuations $\sim 10^{-21}$
Gauss, associated to some early phase transition prior to the EWPT.
The interplay between the resulting polarization effects of the early
Universe plasma and the long range parity violating gauge interaction
present in the Standard Model subsequently amplifies the seed field
till an epoch close to the EWPT time, after which its evolution is
described by standard perturbative physics, free from uncertainties
from the EWPT epoch.

It is well-known that, at the high-temperature symmetric phase of the
Standard Model (SM) all gauge bosons acquire a ``magnetic'' mass gap
$\sim g^2T$, except for the Abelian gauge field associated to weak
hypercharge.  Collective plasma effects also modify the dispersion
relations of the quarks and leptons at finite temperature giving them
effective ``chirally-invariant'' masses determined by their gauge
group quadratic Casimirs~\cite{Weldon:1982bn}.

While long-range magnetic-like fields associated to non-Abelian gauge
forces do not exist, the Abelian hypercharge magnetic fields are never
screened and can survive in the plasma for infinitely long times.
The nonconservation of the lepton charge due to the Abelian anomaly
associated with the hypercharge field $Y_{\mu}$ has been suggested to
play a role in generating the observed the baryon asymmetry of the
Universe~\cite{Giovannini:1997eg}.

Here we stress the importance of polarization effects in the early
Universe plasma associated to the parity violating weak hypercharge
interactions well before the EWPT.  We show how, in the presence of a
lepton asymmetry expected to arise in the leptogenesis
scenario~\cite{Fukugita:1986hr}, a pre-existing field $B_0^Y$, for
simplicity assumed to be a large-scale field, will undergo spectacular
growth prior to the EWPT.  Below the critical EWPT temperature the
hypermagnetic fields are converted into standard Maxwellian magnetic
fields. Thus the amplification mechanism can seed the ordinary
magnetic field, helping to account for the observed magnitude of
intergalactic magnetic fields~\cite{1983flma....3.....Z}.

Consider the equations of motion for the hypercharge $Y_{\mu}$-field
in the hot plasma and in the presence of a pre-existing
large-scale hypermagnetic field ${\bf B}_0^Y$, regular on scales
smaller than the horizon size at $T_0 > T\gg T_{EWPT}$. We assume that
this primordial field has a small amplitude $g^{'}B_0^Y\ll T^2$.  For
simplicity, we neglect the Abelian anomaly and assume flat Minkowski
space. The $U(1)_Y$ interaction Lagrangian in the SM with the one Higgs doublet $\varphi^T=(\phi^{(+)}\phi^{(0)})$
is given as,
\begin{eqnarray}
  \label{eq:Lagrangian}
  &&{\cal L}_{int} = \sum_\ell \frac{g^{'}Y^{\mu}}{2}
\left[-\bar{\nu}_{\ell L}\gamma_{\mu}\nu_{\ell L} 
-\bar{\ell}_L\gamma_{\mu}\ell_L 
- 2\bar{\ell}_R\gamma_{\mu}\ell_R \right] +\nonumber\\&&
+\sum_i^N\frac{g^{'}Y^{\mu}}{2}\left[\frac{1}{3}\bar{U}_{iL}\gamma_{\mu}U_{iL}+\frac{1}{3}\bar{D}_{iL}\gamma_{\mu}D_{iL} +\frac{4}{3}\bar{U}_{iR}\gamma_{\mu}U_{iR} -\frac{2}{3}\bar{D}_{iR}\gamma_{\mu}D_{iR}\right] +\nonumber\\&&+
i\frac{g^{'}Y^{\mu}}{2}\left[\varphi^+{D}_{\mu}\varphi -({D}_{\mu}\varphi^+)\varphi\right].
\end{eqnarray}
For the assumed external seed hypermagnetic field ${\bf
  B}_0^Y=\nabla\times {\bf Y}^{(0)}=(0,0, B_0^Y)$, this leads to the
following Dirac equations for massless charged leptons
($\ell=e,~\mu,~\tau$ ), neutrinos and quarks ($U_i=u,c,t$, $D_i=d,s,b$)
$$ \left[\hat{p}
  -f^{(a)}(g^{'})\hat{Y}^{(0)}\right]{\Psi^{(a)}}=0,$$ where
$a=\ell_L, \ell_{R}, \nu_\ell, U_L, D_L, U_R, D_R$ and
$f^{(a)}(g^{'})$ denote the corresponding SM couplings:
$f_R(g^{'})=-g^{'}$ for right-handed charged leptons,
$f_L(g^{'})=-g^{'}/2$ for left-handed charged leptons and neutrinos,
$f^{(U,D)}_L(g^{'})=g^{'}/6$ and $f^{(U)}_R(g^{'})=2g^{'}/3$,
$f^{(D)}_R(g^{'})=-g^{'}/3$ for left-handed quarks and right-handed
quarks respectively.

The resulting Landau spectrum in the mean hypercharge magnetic field
takes the form $\varepsilon (p_z,n,\lambda)=\sqrt{p_z^2+\mid
  f_{L,R}(g^{'})\mid B^{Y}_0(2n+1 \mp \lambda)}$, where the upper sign
applies to particles and the lower one to antiparticles, irrespective of
their chiralities, $\lambda=\pm 1$.  Within the small-field
approximation $g^{'}B^{Y}_0\ll T^2$ we have
\begin{equation}
\label{eq:spectrum}
\varepsilon (p,\lambda)=p + \mid f_{L,R}(g^{'})\mid B^{Y}_0\lambda/2p
\end{equation} 
with $p=\sqrt{p_z^2+p_{\perp}^2}$ and $p_{\perp}^2=\mid
f_{L,R}(g^{'})\mid B^{Y}_0(2n+1)$. Note that such paramagnetic term in
Eq. (\ref{eq:spectrum}) coming from the spin of fermions is absent for
Higgs bosons.

The above spectrum leads to the unique definition of the equilibrium
distribution functions in the multi-particle approach. Up to the
Abelian anomaly lepton (and quark) numbers are conserved through the
continuity equations $\partial j_{\mu}^{(a)}/\partial x_{\mu}=0$ for
$a=\ell_L, \ell_{R}, \nu_\ell, U_{L,R}, D_{L,R}$.  Therefore we may
define chemical potentials $\mu^{(a)}$ for the corresponding
equilibrium distribution functions.  From Eq.~ (\ref{eq:spectrum}) we
easily find the equilibrium one-particle density matrix for massless
particles $$
f^{(a,\bar{a})}_{\lambda^{'}\lambda}=\frac{\delta_{\lambda^{'}\lambda}}{\exp[(\varepsilon
  (p_z,n,\lambda)\mp\mu_a)/T] +1}$$ which is approximated in the
quasi-classical limit, by
\begin{equation}
\label{eq:matrix}
f^{(a,\bar{a})}_{\lambda^{'}\lambda}\simeq\frac{\delta_{\lambda^{'}\lambda}}{2}f^{(a,\bar{a})}_0(p) + \frac{\sigma_{\lambda^{'}\lambda}^j}{2}S^{(a,\bar{a})j}_0(p),
\end{equation}
with the lower sign for chemical potentials corresponding to
antiparticles. 
Here $f^{(a,\bar{a})}(p)=[{e^{(p \mp \mu_{a})/T}+1}]^{-1}$ are the Fermi
distributions of particles (antiparticles) which for $\mu_a/T\ll 1$
correspond to the Lorentz-invariant densities $n^{(a,\bar{a})}$,
\begin{eqnarray}
\label{eq:density}
&&n^{(a,\bar{a})}=\int \frac{d^3p}{(2\pi)^3}\frac{1}{\exp ([p\mp \mu_a]/T) +1}\approx \nonumber\\&&\approx n_{eq}\left[1 \pm \frac{\pi^2}{9\zeta (3)}\left(\frac{\mu_a}{T}\right) + O\left(\left(\frac{\mu_a}{T}\right)^2\right)\right].
\end{eqnarray}

Here $n_{eq}=3\zeta (3)T^3/4\pi^2$ is the equilibrium lepton density
when $\mu_a=0$; $\zeta (3)\approx 1.202$ is the Riemann function
value.
The second term in Eq.~(\ref{eq:matrix}) includes Pauli matrices
multiplied by the mean spin vector,
\begin{equation}
\label{eq:spin}
{\bf  S}^{(a,\bar{a})}_0(p)=
-\frac{\mid f_a(g^{'})\mid {\bf B}^{Y}_0}{2p}\frac{{\rm d}f^{(a,\bar{a})}_0(p)}{{\rm d}p}=
\hat{\bf b}_0S^{(a,\bar{a})}_0(p),
\end{equation}
which is the equilibrium spin distribution function. Here $\hat{\bf
  b}_0={\bf B}_0^{Y}/B_0^Y$ is the unit vector along the mean
hypermagnetic field.  Notice that equilibrium spin distribution values
$S^{(a,\bar{a})}_0(p)$ define the densities of fermions populating the
main Landau level, $n=0$:
\begin{eqnarray}
\label{eq:Landau}
&&n^{(a, \bar{a})}_0=\int \frac{d^3p}{(2\pi)^3}S^{(a,\bar{a})}_0(p)\\ \nonumber
%=\frac{\mid f_a(g^{'})\mid B_0^Y}{4\pi^2}\int_0^{\infty}f^{(a,\bar{a})}_0(p)dp=\nonumber\\
&&=\frac{\mid f_a(g^{'})\mid B_0^YT\ln 2}{4\pi^2}\left[1 \pm \frac{\mu_a}{2T\ln 2}+ O\left(\left(\frac{\mu_a}{T}\right)^2\right)\right].
\end{eqnarray}
Note also that the subindex $\lambda$ in the density matrix
corresponds to the spin projection on the hypermagnetic field,
$(\sigma_z)_{\lambda^{'}\lambda}=\lambda \delta_{\lambda^{'}\lambda}$.
Together with chirality $\gamma_5\Psi_{e_{R,L}}=\pm \Psi_{e_{R,L}}$ it
is a good quantum number since $[\gamma_5,\Sigma_z]=0$.

Note that the density asymmetry for massless fermions coming from
Eq. (\ref{eq:density}) , $n^{(a)}-n^{(\bar{a})}=\mu_aT^2/6$, is one
half that of the Higgs bosons ($\mu_a T^2/3$). Taking also into
account the equilibrium conditions for the chemical potentials of the
plasma components (using plasma neutrality $<Q>=0$, $<Q_3>=\mu_W=0$
\cite{Harvey:1990qw}) and normalizing the hypercharge $Y$ on $T^2/6$
where $Y$ directly reads from the Lagrangian Eq. (\ref{eq:Lagrangian})
as \footnote{The same result follows from the definition of
  hypercharge $Y$ in Eq. (8) in paper \cite{Khlebnikov}.},
\begin{eqnarray}\label{hypercharge}
&&Y=-\sum_l(\mu_{\nu_lL} + \mu_L^{l}) -2\sum_l\mu_R^{l} + N[\mu_{uL}+\mu_{dL} + 4\mu_{uR} -2\mu_{dR}] + 2(\mu_++\mu_0)=\nonumber\\&&=2(Q-Q_3)=2\left[-2\sum_l\mu_L^{l} + 6\mu_{uL} + 14\mu_0\right]=0,
\end{eqnarray}
one finds the chemical potential of the neutral Higgs boson $\mu_0$ , \begin{equation}\label{Higgs}
\mu_0=\frac{\sum_l\mu_L^l-3\mu_{uL}}{7}~.
\end{equation}
Using the sphaleron equilibrium condition valid above EWPT,
\begin{equation}
  \label{eq:sphal-eq}
\sum_l\mu_L^l=-9\mu_{uL}  
\end{equation}
and the expressions for the baryon and lepton numbers in terms of
chemical potentials (given in Eq. (8) in Ref.~\cite{Harvey:1990qw})
one can see that $\mu_0$ vanishes only in the exceptional case $B=L=0$ (see, e.g.
in \cite{Kainulainen}).

Note that in the first line of Eq. (\ref{hypercharge}) we have used
$\mu_+=-\mu_-=\mu_0$. This follows from the equilibrium relation
$\mu_W=\mu_ -+\mu_0$ (implied by conversions $W^-\leftrightarrow
\phi^- + \phi^0$) and the neutrality condition for the isospin
component $Q_3\sim \mu_W=0$.
We also took into account the common color factor of quarks and
assumed that quark asymmetries are degenerate in flavor for all
left-handed up-quark fields and all right-handed down-quark fields
respectively (note that we do not assume such degeneracy for charged
leptons and neutrinos).

One can easily write down Maxwell-like equations of motion for the
Abelian $Y_{\mu}$-field in the equilibrium plasma. These involve
thermal averages of the various four-current fermion densities, e.~g.
for leptons $<\bar{\ell}_a\gamma_{\mu}\ell_a>$, given as differences
of Wigner distribution functions for particles and antiparticles,
$\delta f^{(a)}_{\lambda^{'}\lambda}({\bf p}, {\bf
  x},t)=f^{(a)}_{\lambda^{'}\lambda}({\bf p}, {\bf
  x},t)-f^{(\bar{a})}_{\lambda^{'}\lambda}({\bf p}, {\bf x},t)$.
Substituting the general distribution function,
\begin{equation}
\label{eq:Wigner}
f^{(a)}_{\lambda^{'}\lambda}({\bf p}, 
{\bf x}, t)=\frac{\delta_{\lambda^{'}\lambda}}{2}f^{(a)}({\bf p}, {\bf x}, t) + 
\frac{(\sigma_i)_{\lambda^{'}\lambda}}{2}S^{(a)}_i({\bf p}, {\bf x}, t),
\end{equation} 
and summing over spin variables $\lambda$, $\lambda^{'}$ one can
recast the Poisson and Maxwell-like equations as
\begin{eqnarray}
\label{eq:fieldequations}
&&\nabla\cdot{\bf B}_Y=0,~~~\nabla\cdot{\bf E}_Y=4\pi\left[J_0^Y({\bf x},t) + 
J^Y_{05}({\bf x},t)\right],\nonumber\\&&
\frac{\partial {\bf B}_Y}{\partial t}= - \nabla\times {\bf E}_Y,\nonumber\\
&&-\frac{\partial {\bf E}_Y}{\partial t} + 
\nabla\times {\bf B}_Y=4\pi\left[{\bf J}^Y({\bf x},t) + {\bf J}^Y_5({\bf x},t)\right],
\end{eqnarray}
These describe the equilibrium plasma at $T\gg T_{EWPT}$ and differ
from the familiar Maxwell equations by the presence of the
pseudovector current $J^Y_{\mu5}$ associated to the parity violating
$Y_{\mu}$ interactions.

The vector current is a sum  $J_{\mu}^Y = \sum_\ell J_{\ell
  \mu}^Y + 3NJ^Y_{(q)\mu} + J_{(\varphi) \mu}^Y$ where each term is given in terms of differences of current
asymmetries $$\delta j_{\mu}^{(a)}= j^{(a)}_{\mu} -
j_{\mu}^{(\bar{a})}=\int \frac{d^3p}{(2\pi)^3} \frac{p_\mu}{p} \delta
f^{(a)} ({\bf p}, {\bf x}, t)$$ in the form following from the interaction Lagrangian (\ref{eq:Lagrangian}),
\begin{eqnarray}
\label{eq:vector1}
  &&J_{\ell \mu}^Y({\bf x},t)=
 -\frac{g^{'}}{4}\left[2\delta j_{\mu}^{\ell_R}({\bf x},t) 
 +\delta j_{\mu}^{\ell_L}({\bf x},t)
 +\delta j_{\mu}^{\nu_{\ell L}}({\bf x},t)\right],\nonumber\\&&
 J_{(q)\mu}^Y({\bf x},t)=\frac{g^{'}}{12}\left[\left(\delta j_{\mu}^{(uL)}({\bf x},t)+\delta j_{\mu}^{(dL)}({\bf x},t)\right)+4\delta j_{\mu}^{(uR)}({\bf x},t) -2\delta j_{\mu}^{(dR)}({\bf x},t)\right],\nonumber\\&&
 J_{(\varphi) \mu}^Y({\bf x},t)=\frac{g^{'}}{2}\delta j_{\mu}^{(\varphi)}({\bf x},t),
\end{eqnarray}
involving partial current asymmetries given as
\begin{equation}
\label{eq:currentasymmetry}
\delta j^{(a)}_{\mu}=\{(n^{(a)}- n^{(\bar{a})})\gamma;~~~n_{eq}
\gamma ({\bf V}^{(a)} - {\bf V}^{(\bar{a})})\},
\end{equation}
where $n^{(a)}\equiv n^{(a)}({\bf x},t)=j_{\mu}^{(a)}({\bf
  x},t)u^{\mu}({\bf x},t)$ are the densities defined in Eq.
(\ref{eq:density}) which coincide with those calculated in the medium
rest frame, ${\bf V}=0$, $\gamma=(1 - V^2)^{-1/2}=1$.  Summing the
individual contributions in Eq.~(\ref{eq:vector1}) one obtains the
hypercharge vector current density $J_{\mu}^Y$ as
\begin{eqnarray}
\label{eq:vectorcurrent}
&&J_{0}^Y=-\gamma n_{eq}\left(\frac{2\pi^2}{9\zeta(3)}\right)\left(\frac{g^{'}}{4T}\right)\left[-2\sum_l\mu_L^{l} + 6\mu_{uL} + 14\mu_0\right],\nonumber\\&&
{\bf J}^Y=\sum_a\frac{f^{(a)}(g^{'})}{2}\gamma n_{eq}\left[{\bf V}^{(a)} - {\bf V}^{(\bar{a})}\right].
\end{eqnarray}

As seen from Eq. (\ref{hypercharge}) hypercharge neutrality of the
plasma $<Y>=0$ implies the vanishing the component $J_0^Y=0$ at large
scales exceeding the Debye radius, $r\gg r_{D}$, resulting in the
relation between chemical potentials Eq. (\ref{Higgs}).

We now turn to the axial current.  Using Eq.~(\ref{eq:Wigner}) one can
express the axial vector current $J_{\mu 5}^Y({\bf x},t)$ in terms of
differences of spin distribution functions entering the axial
four-vectors $\delta A_{\mu}^{(a)}({\bf p}, {\bf
  x},t)=A_{\mu}^{(a)}({\bf p}, {\bf x},t)-A_{\mu}^{(\bar{a})}({\bf p},
{\bf x},t)$ with $A_{\mu}^{(a)}({\bf p}, {\bf x},t )=\{{\bf p}\cdot
{\bf S}^{(a)}({\bf p}, {\bf x},t));({\bf p}\cdot {\bf S}^{(a)}({\bf
  p}, {\bf x},t)){\bf p}/p\}$~\footnote{This is the massless limit of
  the four-vector spin distribution that generalizes the
  Pauli-Lubanski spin vector in the multi-particle approach
  \cite{Semikoz:1987cv}, $A_{\mu}^{(a)}({\bf p}, {\bf
    x},t)=m_{a}\times a_{\mu}=\lim_{m_{a}\to 0}m_{a}\times[{\bf p}{\bf
    S}^{(a)}/m_{a};{\bf S}^{(a)} +({\bf p}{\bf S}^{(a)}){\bf
    p}/m_{a}(\varepsilon_p + m_{a})]$, where ${\bf S}^{(a)}={\bf
    S}^{(a)}({\bf p}, {\bf x},t)$.  Note that the standard
  Lorenz-invariant normalization $a_{\mu}a^{\mu}=-{\bf S}^2\neq 0$ is
  replaced by $A_{\mu}^{(a)}A^{\mu (a)}=0$ in the massless fermion
  case.}.  
  
The total pseudovector current $J_{\mu 5}^Y=\sum_lJ^Y_{l\mu 5}({\bf
  x},t) + 3NJ_{q\mu 5}^Y$ is given in terms of partial contributions
of leptons and quarks:
\begin{eqnarray}
\label{eq:axialvector}
&&J_{\ell \mu5}^Y({\bf x},t)=
-\frac{g^{'}}{2}\int\frac{d^3p}{p(2\pi)^3}\delta A_{\mu}^{(\ell R)}({\bf p}, {\bf x},t)
+\frac{g^{'}}{4}\int\frac{d^3p}{p(2\pi)^3}\delta A_{\mu}^{(\ell L)}({\bf p}, {\bf x},t)+ \nonumber\\&&+\frac{g^{'}}{4}\int\frac{d^3p}{p(2\pi)^3}\delta A_{\mu}^{(\nu_\ell)}({\bf p}, {\bf x},t)~,
\nonumber\\&&J^Y_{q\mu 5}=-\frac{g^{'}}{12}\int\frac{d^3p}{p(2\pi)^3}\delta A_{\mu}^{(uL)}({\bf p}, {\bf x},t)-\frac{g^{'}}{12}\int\frac{d^3p}{p(2\pi)^3}\delta A_{\mu}^{(dL)}({\bf p}, {\bf x},t) +\nonumber\\&&+
\frac{g^{'}}{3}\int\frac{d^3p}{p(2\pi)^3}\delta A_{\mu}^{(uR)}({\bf p}, {\bf x},t)-\frac{g^{'}}{6}\int\frac{d^3p}{p(2\pi)^3}\delta A_{\mu}^{(dR)}({\bf p}, {\bf x},t)~.
\end{eqnarray}
Substituting the equilibrium spin distribution Eq.~(\ref{eq:spin})
into Eq.~(\ref{eq:axialvector}) we immediately get $J_{05}=0$ in the
rest frame of the medium as a whole, ${\bf V}=0$, $\gamma=1$.
Thus, under the hypercharge neutrality condition in
Eq.~(\ref{hypercharge}), one has that $J_0^Y=J_{05}^Y=0$, so that the
Poisson equation takes the standard form, $\nabla\cdot {\bf E}_Y=0$,
in agreement with Ref.~\cite{Giovannini:1997eg}. (Note, however, that
small scale fluctuations described by non-equilibrium distribution
functions $\delta f^{(a)}({\bf p}, {\bf x},t)$, $\delta {\bf
  S}^{(a)}({\bf p}, {\bf x},t)$ could lead to $\delta J_0^Y\neq 0$,
$\delta J_{05}\neq 0$).

In contrast, each pseudovector 3-current ${\bf J}_{\ell5}^Y$ or ${\bf J}^Y_{q5}$ in Eq.
(\ref{eq:axialvector}) is nonzero even in equilibrium,
\begin{eqnarray}\label{pseudovector}
&&({\bf J}^Y_{\ell5})_{eq}= 
-\frac{g^{'}}{2}[{\bf j}^{\ell R}_5 -{\bf j}^{\bar{\ell} R}_5] 
+\frac{g^{'}}{4}[{\bf j}^{\ell L}_5 -{\bf j}^{\bar{\ell} L}_5]  
+\frac{g^{'}}{4}[{\bf j}^{\nu_\ell }_5 -{\bf j}^{\bar{\nu}_\ell }_5], \nonumber \\
&&({\bf J}^Y_{q5})_{eq}=-\frac{g^{'}}{12}\left[{\bf j}_5^{uL}-{\bf j}_5^{\bar{u}L}\right]-\frac{g^{'}}{12}\left[{\bf j}_5^{dL}-{\bf j}_5^{\bar{d}L}\right]+\frac{g^{'}}{3}\left[{\bf j}_5^{uR}-{\bf j}_5^{\bar{u}R}\right]-\frac{g^{'}}{6}\left[{\bf j}_5^{dR}-{\bf j}_5^{\bar{d}R}\right].
\end{eqnarray}
Substituting the equilibrium spin distributions in Eq.~(\ref{eq:spin})
and using again the hypercharge neutrality condition and the sphaleron
equilibrium condition Eq.~(\ref{eq:sphal-eq}) we get, after summing
over leptons $\sum_\ell$ and quarks,
\begin{equation}
\label{eq:current5}
({\bf J}^Y_{5})_{eq} =  \frac{g^{'2}}{96\pi^2}\left[-2\sum_l\mu_L^l + 10\mu_{uL} + 32\mu_0 \right]{\bf B}_0^Y=\frac{47}{1512\pi^2}g^{'2}\mu_{\nu}{\bf B}_0^Y~,
\end{equation}
which leaves then only one independent lepton asymmetry which we take
as that of neutrinos,
$\sum_l\mu_L^l=\sum_l\mu_{\nu_{lL}}=\mu_{\nu}$. Notice that we have
used here Eqs. (\ref{Higgs}), (\ref{eq:sphal-eq}) for the chemical
potentials $\mu_0$ and $\mu_{uL}$. 

Thus, the system of Magneto-Hydro-Dynamics (MHD) equations for
hyperelectromagnetic fields in Eq.~(\ref{eq:fieldequations}) finally
gets the form:
\begin{eqnarray}
\label{eq:fieldequations2}
&&\nabla\cdot{\bf B}_Y=0,~~~~~~~~~\nabla\cdot{\bf E}_Y= 0,\nonumber\\&&
\frac{\partial {\bf B}_Y}{\partial t}= - \nabla\times {\bf E}_Y,\nonumber\\&&-\frac{\partial {\bf E}_Y}{\partial t} + \nabla\times {\bf B}_Y=4\pi{\bf J}^Y 
+\frac{47}{378}\times \frac{g^{{'}2}\mu_\nu}{\pi}{\bf B}_0^Y.
\end{eqnarray}

Averaging the total field ${\bf B}_Y={\bf B}_0^Y + {\bf b}_Y({\bf
  x},t)$ over random small-scales, $<{\bf b}_Y>=0$, we can rewrite
Eq.~(\ref{eq:fieldequations2}) as an MHD system for mean hypermagnetic
fields ${\bf B}_Y={\bf B}_0^Y$ completed by the Ohm equation. In the
rest frame ${\bf V}=0$ of the isotropic early Universe plasma we are
considering the Ohm equation reduces to
\begin{equation}
\label{eq:Ohm}
{\bf J}^Y=\sigma_{cond}{\bf E}_Y.
\end{equation}
One sees that the MHD equations derived here in the standard
Weinberg-Salam model qualitatively coincide with what one obtains
using the anomaly term in the interaction Lagrangian
\cite{Joyce:1997uy,Redlich:1984md}.

By combining the Ohm law Eq.~(\ref{eq:Ohm}) and the last Maxwell-like
equation in Eq.~(\ref{eq:fieldequations2}) then using 
$\partial {\bf B}_Y/\partial t=-\nabla\times {\bf E}_Y$, we can write the 
Faraday equation describing the so-called $\alpha^2$-dynamo
\cite{1983flma....3.....Z} of hypermagnetic field as
\begin{equation}
\label{eq:Faraday}
\frac{\partial {\bf B}_Y}{\partial t}= 
\nabla\times \alpha {\bf B} + \eta\nabla^2{\bf B}_Y,
\end{equation} 
where $\eta=(4\pi\sigma_{cond})^{-1}$ is the magnetic diffusion coefficient and we neglect, as
usual in MHD, the displacement current $\partial {\bf E}_Y/\partial t$
and use the rest frame condition ${\bf V}=0$. The parameter $\alpha$
is the hypermagnetic helicity coefficient given as
\begin{equation}
\label{eq:alpha}
\alpha=\frac{47g^{{'}2}\mu_\nu}{1512\pi^2\sigma_{cond}}
\end{equation}
and plays crucial role in the evolution of the hypermagnetic field.
We can solve Eq.~(\ref{eq:Faraday}) through Fourier harmonics as
  ${\bf B}_Y({\bf x},t)=\int (d^3k/(2\pi)^3{\bf B_Y}({\bf
    k},t)e^{i{\bf k}{\bf x}}$ where $B_Y(k,t)$ is expressed as
\begin{equation}
\label{eq:growth0}
B_Y(k,t)=B_0^Y\exp 
\left[\int_{t_{0}}^t[\alpha (t^{'})k-\eta (t^{'})k^2)]{\rm d}t^{'}\right]. 
\end{equation}
For $0<k<\alpha/\eta$, or correspondingly correlation length scales
$\eta/\alpha<\Lambda<\infty$ such field gets exponentially amplified,
but differently for different scales $\Lambda$. E.g. for the Fourier
mode $k=\alpha/2\eta$ (or $\Lambda\sim 2\eta/\alpha$) one gets the
maximum amplification $\gamma=\alpha k - \eta
k^2=\alpha^2/4\eta$~\cite{1983flma....3.....Z,Semikoz:2003qt}
$$
B_Y(t)=B_0^Y\exp \left[\int_{t_0}^{t}\frac{\alpha^2(t^{'})}{4\eta (t^{'})}{\rm d}t^{'}\right]
$$
or
\begin{equation}
\label{eq:growth}
B_Y(x) = B_0^Y\exp\left[32\int_x^{x_0}\frac{{\rm d}x^{'}}{x^{'2}}\left(\frac{\xi_{\nu}(x^{'})}{0.001}\right)^2\right]
\end{equation}
where we introduced the new variable $x=T/T_{EWPT}$ and $B_0^Y$ is the
assumed initial amplitude of the hypermagnetic field at $T_0\gg
T_{EWPT}$. In the second equality we substituted $\eta=(4\pi
137T\cos^2\theta_W)^{-1}$, the analogue of the magnetic diffusion
coefficent taking into account the change from the standard QED fine
structure constant ($e^2\sim 137^{-1}$) to the analogous hypercharge
one, with $e^2\to g^{'2}=e^2/\cos^2\theta_W$, where
$\sin^2\theta_W=0.23$ is the electroweak mixing parameter.
We have also used the appropriate cosmological time-temperature
relation $t=[3.84\times 10^{21}(T/{\rm MeV})^{-2}/\sqrt{g^*}]~{\rm
  MeV}^{-1}$ with the number of relativistic degrees of freedom
$g^{*}\sim 100$.
One sees that, even for small values of the lepton asymmetry
$\xi_{\nu}$ one obtains a very strong amplification of the seed
hypermagnetic field $B_0^Y$ at this scale.

Unfortunately, the unknown dependence of $\xi_{\nu}(x)$ for
$x=T/T_{EWPT}>1$ prevents us from a reliable numerical estimate for the
amplitude $B_Y(x)$. Moreover, since the evolution of $\xi_\nu$ depends
on nonperturbative physics, it is not easy at this stage to confront
it with the primordial nucleosynthesis bounds~\cite{Dolgov:2002ab}.
However, in order to survive against ohmic dissipation due to finite
conductivity $\eta=(4\pi\sigma_{cond})^{-1}$ we should have
$\Lambda>l_{diff}$, where $l_{diff}=\sqrt{\eta l_H}$ is the diffusion
length. This leads to an upper bound on $\xi_{\nu}(x)=\mu_{\nu}(T)/T$,
\begin{equation}\label{asymmetrybound}
\frac{\xi_{\nu}(x)}{0.001}< A(\Lambda)\sqrt{x}~,~~~~~~~~x\geq 1,
\end{equation}
which explicitly depends on the chosen scale $\Lambda$.  Hence for the
mode $\Lambda_1=2\eta/\alpha$ one gets $A=0.23$ so that the
amplification factor is $\sim 32$ as seen in Eq. (\ref{eq:growth}), while for a larger scale, say
$\Lambda_2=16\eta/\alpha$, one gets $A=1.83$ with a reduced growth
factor $\sim 7.4$ coming from the general solution in Eq. (\ref{eq:growth0}): 
\begin{equation}\label{largescale}
B_Y(t)=B_0^Y\exp \left[\frac{15}{256}\int_{t_0}^{t}\frac{\alpha^2(t^{'})}{\eta (t^{'})}{\rm d}t^{'}\right]=B_0^Y\exp\left[7.4\int_x^{x_0}\frac{{\rm d}x^{'}}{x^{'2}}\left(\frac{\xi_{\nu}(x^{'})}{0.001}\right)^2\right]~.
\end{equation}
In any case one can have very strong amplification
even for larger scales for which the factor $A(\Lambda)$ is
bigger.

In contrast to the mechanism suggested in Refs.~\cite{Joyce:1997uy}
and \cite{Giovannini:1997eg} ours does not rely on the Chern-Simons
anomaly term in the SM Lagrangian~\cite{Redlich:1984md}.  The presence
of the anomaly acting at the later EWPT epoch, could play an important
role in the subsequent evolution of the lepton asymmetry produced by
the parity violating hypercharge interaction.
However, by then the asymmetry has already induced the strong
Maxwellian magnetic fields which no longer convert to leptons, as
these carry no anomaly, their evolution being described by standard
MHD equations. 
Here we do not study the EWPT conversion of hypercharge field to the
Maxwellian magnetic field ${\bf B}$. However we note that the seed
value $B_Y\sim 0.3T^2< T_{EWPT}^2\sim 10^{24}~Gauss$ can be easily
reached through our Eqs. (\ref{eq:growth}), (\ref{largescale}). This provides a strongly
first order EWPT that, in turn, allows to avoid the sphaleron
constraint for baryogenesis within the Standard Model
\cite{Elmfors}. It is also important to note that our bound on the
neutrino asymmetry in Eq. (\ref{asymmetrybound}) provides the large
scale of the mean hypermagnetic field $\Lambda\simeq L_0$ in Eq. (1) of ref. 
\cite{Elmfors},
so that bubble formation during EWPT takes place in the background of
essentially constant field~\cite{Elmfors}.
The subsequent evolution of the Maxwellian field proceeds through the
inverse cascade~\cite{Brandenburg:1996fc}. 

Let us now comment on the physical interpretation of the new magnetic
helicity term.  The original seed field $B_0^Y$ polarizes the fermions
and antifermions (including neutrinos) propagating along the field in
the main Landau level, $n=0$. This polarization effect causes fermions
and antifermions to move in opposite directions with a relative drift
velocity proportional to the lepton asymmetry. The existence of a
basic parity violating hypercharge interaction in the SM induces a new
term in the hypermagnetic field in Eq.~(\ref{eq:Faraday})
$\nabla\times \alpha {\bf B}_Y$ which winds around the rectilinear
pseudovector hypercharge current ${\bf J}_5$ parallel to ${\bf B}_Y$.
This term amplifies the seed hypermagnetic field $B_0^Y$ according to
Eqs.~(\ref{eq:growth0}) and (\ref{eq:growth}).
It is interesting also to consider in detail the hypermagnetic
helicity $H=\int {\rm d}^3x{\bf Y}{\bf B}_Y$ and how it becomes the
magnetic one below the EWPT~\cite{Semikoz:2004rr}.

In summary, while we are far from having a complete and fully
quantitative picture for the origin of intergalactic magnetic fields,
we think that the amplification mechanism described here could play an
important role towards the goal of accounting for the observed
intergalactic magnetic fields from first principles.

\acknowledgments Work supported by MEC grant FPA2005-01269, by EC
Contracts RTN MRTN-CT-2004-503369 and ILIAS/N6 RII3-CT-2004-506222 and
ACOMP07/270 of Generalitat Valenciana.  V. B. S.  thanks support from
CSIC-RAS. We thank discussions with Dimitri Sokoloff.

\end{document}